\documentclass[a4paper,12pt]{article}

\usepackage[dvips]{graphicx}
\usepackage{epsfig}

\begin{document}

\author{C. Barrab\`es\thanks{E-mail : barrabes@celfi.phys.univ-tours.fr}\\
\small Laboratoire de Math\'ematiques et Physique Th\'eorique,\\
\small CNRS/UMR 7350, F\'ed\'eration Denis Poisson CNRS/FR2964, \\
\small Universit\'e F. Rabelais, 37200 TOURS,
France
\\\small and \\P. A. Hogan\thanks{E-mail : peter.hogan@ucd.ie}\\
\small School of Physics,\\ \small  University College Dublin, Belfield, Dublin 4, Ireland}

\title{Collision of Shock Waves in Einstein-Maxwell Theory with a Cosmological Constant: A Special Solution}

\date{\small PACS numbers: 04.40.Nr, 04.30.Nk}
\maketitle

\begin{abstract}Post--collision space--times of the Cartesian product form $M'\times M''$, where $M'$ and $M''$ are 2--dimensional manifolds, are 
known with $M'$ and $M''$ having constant curvatures of equal and opposite sign (for the collision of electromagnetic shock waves) or of the same sign (for 
the collision of gravitational shock waves). We construct here a new explicit post--collision solution of the Einstein--Maxwell vacuum field equations with a cosmological 
constant for which $M'$ has constant (non--zero) curvature and $M''$ has zero curvature.
\end{abstract}
\thispagestyle{empty}
\newpage

\section{Introduction}\indent
The space--time following the head--on collision of two homogeneous, plane, electromagnetic shock waves was found by Bell and Szekeres \cite{BS} and is a solution of the vacuum Einstein--Maxwell 
field equations. The metric tensor is that of a Cartesian product  of two 2--dimensional manifolds of equal but opposite sign constant curvatures and is the Bertotti--Robinson (\cite{B}, \cite{R}) space--time. Recently 
we have shown (\cite{BH2}, \cite{BH3}) that the Nariai--Bertotti (\cite{B}, \cite{N}) space--time, with metric that of a Cartesian product of two 2--dimensional manifolds of equal constant curvatures, coincides with the 
space--time following the head--on collision of two homogeneous, plane, gravitational shock waves and is a solution of Einstein's vacuum field equations with a cosmological constant. We construct here 
a metric for a space--time which is a Cartesian product of two 2--dimensional manifolds, one having \emph{non--zero constant curvature} and one having \emph{zero curvature}, and show that the metric is: (I) that of 
the post collision region of space--time following the head--on collision of two plane light--like signals each consisting of combined gravitational and electromagnetic shock waves, one signal specified 
by a real parameter $a$ and the second signal specified by a real parameter $b$ and (II) is a solution of the vacuum Einstein--Maxwell field equations with a cosmological 
constant $\Lambda=2ab$. The appearance of a cosmological constant term on the left hand side of the Einstein field equations is 
equivalent to the appearance of an energy--momentum--stress tensor for a perfect fluid for which the sum of the matter proper density and the isotropic pressure vanishes. Thus our space--time consists of 
an anti--collision region which is vacuum and a post--collision region which is non--vacuum in this sense. Vacuum and non--vacuum regions of space--time are familiar from solving the field equations 
for so--called interior and exterior solutions. 

\section{Cartesian Product Space--Time}
\setcounter{equation}0\noindent
We consider a pseudo--Riemannian space--time $M$ of the form $M=M'\times M''$ where $M'$ is a 2--dimensional manifold of non--zero constant curvature and $M''$ is a 2--dimensional flat manifold. So that 
the 4--dimensional manifold $M$ has the correct Lorentzian signature we consider the two cases in which (i) $M'$ is pseudo--Riemannian and $M''$ is Riemannian and (ii) $M'$ is Riemannian and $M''$ is 
pseudo--Riemannian. In either case take $\xi, x$ as local coordinates on $M'$ and $\eta, y$ as local coordinates on $M''$. With $a, b$ real constants we take $ab<0$ for case (i) and write the 
line element of $M$ as 
\begin{equation}\label{1}
ds^2=d\xi^2-\cos^2(2\sqrt{-ab}\xi)\,dx^2-d\eta^2-dy^2\ .\end{equation}In terms of the basis 1--forms $\vartheta^1=d\xi$ and $\vartheta^2=\cos(2\sqrt{-ab}\xi)\,dx$ the single non--vanishing Riemann curvature 
tensor component on the dyad defined by this basis, for the manifold $M'$, is 
\begin{equation}\label{2}
R_{1212}=4ab\ ,\end{equation}indicating that the pseudo--Riemannian manifold $M'$ has non--zero constant Riemannian curvature (see, for example, \cite{SS}) $-4ab>0$. Clearly the manifold $M''$ is Riemannian and flat. For 
case (ii) we take $ab>0$ and write the line element of $M$ as
\begin{equation}\label{3}
ds^2=-d\xi^2-\cos^2(2\sqrt{ab}\xi)\,dx^2+d\eta^2-dy^2\ .\end{equation}Now $M'$ is Riemannian. In terms of the basis 1--forms $\vartheta^1=d\xi$ and 
$\vartheta^2=\cos(2\sqrt{ab}\xi)\,dx$ the non--vanishing component of the Riemann curvature tensor for $M'$, on the dyad defined by the basis 1--forms, is 
\begin{equation}\label{4}
R_{1212}=-4ab\ ,\end{equation}indicating that the Riemannian manifold $M'$ has non--zero Gaussian curvature $K=-R_{1212}=4ab>0$. In this case the manifold $M''$ is pseudo--Riemannian and flat. Now 
for case (i) make the transformation
\begin{equation}\label{5}
\xi=\frac{au-bv}{\sqrt{-2ab}}\ ,\ \ \eta=\frac{au+bv}{\sqrt{-2ab}}\ ,\end{equation}while for case (ii) make the transformation
\begin{equation}\label{6}
\xi=\frac{au-bv}{\sqrt{2ab}}\ ,\ \ \eta=\frac{au+bv}{\sqrt{2ab}}\ .\end{equation} In both cases the line elements (\ref{1}) and (\ref{3}) become
\begin{equation}\label{7}
ds^2=-\cos^2\{\sqrt{2}(au-bv)\}dx^2-dy^2+2\,du\,dv\ .\end{equation} We can write this line element in the form
\begin{equation}\label{8}
ds^2=-(\vartheta^1)^2-(\vartheta^2)^2+2\,\vartheta^3\vartheta^4=g_{ab}\vartheta^a\vartheta^b\ ,\end{equation}with the basis 1--forms given, for example, by $\vartheta^1=\cos\{\sqrt{2}(au-bv)\}dx,\ 
\vartheta^2=dy,\ \vartheta^3=dv,\ \vartheta^4=du$. Thus the constants $g_{ab}$ are the components of the metric tensor on the half--null tetrad defined via the basis 1--forms. The components $R_{ab}$ of 
the Ricci tensor on this tetrad vanish except for
\begin{equation}\label{9}
R_{11}=-4ab\ ,\ \ \ R_{33}=-2b^2\ ,\ \ \ R_{34}=2ab\ ,\ \ \ R_{44}=-2a^2\ .\end{equation}With
\begin{equation}\label{10}
F=\frac{1}{2}F_{ab}\vartheta^a\wedge\vartheta^b=a\,\vartheta^1\wedge\vartheta^4+b\,\vartheta^3\wedge\vartheta^1\ ,\end{equation}and $\Lambda=2ab$ we have here a solution of the Einstein--Maxwell 
vacuum field equations with a cosmological constant:
\begin{equation}\label{11}
R_{ab}=\Lambda\,g_{ab}+2E_{ab}\ ,\end{equation}and
\begin{equation}\label{12}
dF=0=d{}^*F\ ,\end{equation}where $d$ denotes the exterior derivative, ${}^*F=a\,\vartheta^2\wedge\vartheta^4+b\,\vartheta^2\wedge\vartheta^3$ is the Hodge dual of the Maxwell 2--form (\ref{10}) 
with components $F_{ab}$ on the tetrad given by (\ref{10}) and $E_{ab}=F_{ac}F_b{}^c-\frac{1}{4}g_{ab}\,F_{cd}F^{cd}$ is the electromagnetic energy--momentum tensor. Tetrad indices are raised with 
$g^{ab}$ where $g^{ab}g_{bc}=\delta^a_c$. In Newman--Penrose \cite{NP} notation, the Weyl tensor has components
\begin{equation}\label{13}
\Psi_0=b^2\ ,\ \ \Psi_1=0\ ,\  \ \Psi_2=\frac{1}{3}ab\ ,\ \ \Psi_3=0\ ,\ \ \Psi_4=a^2\ ,\end{equation}which is type D in the Petrov classification and the Maxwell tensor, given by (\ref{10}), has components
\begin{equation}\label{14}
\Phi_0=b\ ,\ \ \Phi_1=0\ ,\ \ \Phi_2=a\ .\end{equation}

\section{Collision of Light--Like Signals}
\setcounter{equation}0\noindent
To demonstrate that the space--time with line element (\ref{7})  and the Maxwell field (\ref{10}) describe the gravitational and electromagnetic fields following the head--on collision of two homogeneous, plane,  
light--like signals, each composed of an electromagnetic shock wave accompanied by a gravitational shock wave, we replace $u, v$ in the argument of the cosine in (\ref{7}) by $u_+=u\vartheta(u), v_+=v\vartheta(v)$ 
where $\vartheta(u)$ is the Heaviside step function which is equal to unity for $u>0$ and is zero for $u<0$ (and similarly for $\vartheta(v)$) so that the line element we now consider reads 
\begin{equation}\label{15}
ds^2=-\cos^2\{\sqrt{2}(au_+-bv_+)\}dx^2-dy^2+2\,du\,dv\ .\end{equation} Writing this line element in the form (\ref{8}) with basis 1--forms now given by $\vartheta^1=\cos\{\sqrt{2}(au_+-bv_+)\}dx,\ 
\vartheta^2=dy,\ \vartheta^3=dv,\ \vartheta^4=du$ we find that the components $R_{ab}$ of the Ricci tensor on the tetrad defined by this basis of 1--forms vanish except for 
\begin{eqnarray}
R_{11}&=&-4ab\,\vartheta(u)\vartheta(v)\ ,\ \ R_{33}=b\sqrt{2}\,\delta(v)\tan(\sqrt{2}\,au_+)-2b^2\vartheta(v)\ ,\nonumber\\
R_{34}&=&2ab\,\vartheta(u)\vartheta(v)\ ,\ \ R_{44}=a\sqrt{2}\,\delta(u)\tan(\sqrt{2}\,bv_+)-2a^2\vartheta(u)\ .\label{16}\end{eqnarray}This Ricci tensor can be written in the form
\begin{equation}\label{17}
R_{ab}=\Lambda\,g_{ab}+2E_{ab}+S_{ab}\ ,\end{equation}with $\Lambda=2\,ab\vartheta(u)\vartheta(v)$, $E_{ab}$ the tetrad components of the electromagnetic energy--momentum tensor calculated with 
the Maxwell field given by the 2--form
\begin{equation}\label{18}
F=b\,\vartheta(v)\vartheta^3\wedge\vartheta^1+a\,\vartheta(u)\vartheta^1\wedge\vartheta^4\ ,\end{equation}and $S_{ab}$ are the components of the surface stress--energy tensor \cite{BH} concentrated 
on the portions of the null hypersurfaces $u=0, v>0$ and $v=0, u>0$ and given by
\begin{equation}\label{19}
S_{ab}=b\sqrt{2}\,\delta(v)\,\tan(\sqrt{2}\,au_+)\delta^3_a\delta^3_b+a\sqrt{2}\,\delta(u)\tan(\sqrt{2}\,bv_+)\delta^4_a\delta^4_b\ .\end{equation}We emphasize that in the post collision 
domain ($u>0, v>0$) the field equations (\ref{17}) can be written in the form
\begin{equation}\label{19'}
R_{ab}-\frac{1}{2}\,g_{ab}\,R=T_{ab}+2\,E_{ab}\ \ \ {\rm with}\ \ \ T_{ab}=-2\,a\,b\,g_{ab}\ ,\end{equation}where $R$ denotes the Ricci scalar.While the term $T_{ab}$ on the right hand side here has 
the form of a cosmological constant term it is equivalent to the energy--momentum--stress tensor for a perfect fluid for which the sum of the matter proper density and the isotropic pressure vanishes.

The Newman--Penrose components of the Maxwell field (\ref{18}) 
are thus
\begin{equation}\label{20}
\Phi_0=b\,\vartheta(v)\ ,\ \ \Phi_1=0\ ,\ \ \Phi_2=a\,\vartheta(u)\ ,\end{equation}while the Newman--Penrose components of the Weyl tensor are
\begin{eqnarray}
\Psi_0&=&-\frac{1}{\sqrt{2}}b\,\delta(v)\tan(\sqrt{2}\,au_+)+b^2\vartheta(v)\ ,\ \ \Psi_1=0\ ,\nonumber\\
\Psi_2&=&\frac{1}{3}ab\,\vartheta(u)\vartheta(v)\ ,\label{21}\\
\Psi_3&=&0\ ,\ \ \Psi_4=-\frac{1}{\sqrt{2}}a\,\delta(u)\,\tan(\sqrt{2}\,bv_+)+a^2\vartheta(u)\ .\nonumber\end{eqnarray}On account of the appearance of the trigonometric functions in 
(\ref{19}) and (\ref{21}) we see that the coordinate $u$ has the range $0\leq u<\pi/2\sqrt{2}\,a$ on $v=0$ and the coordinate $v$ has the range 
$0\leq v<\pi/2\sqrt{2}\,b$ on $u=0$. Such restrictions are also exhibited in the Bell--Szekeres \cite{BS} solution and are discussed in \cite{CH1}.  

We are now in a position to interpret physically what these equations are describing. First 
we consider the region of space--time corresponding to $v<0$. Now $R_{ab}=2E_{ab}$ with $E_{ab}$ constructed from the Maxwell field $a\,\vartheta(u)\vartheta^1\wedge\vartheta^4$. All 
Newman--Penrose components of the Weyl tensor vanish except $\Psi_4=a^2\vartheta(u)$. We have here a solution of the vacuum Einstein--Maxwell field equations for $u>0$ corresponding to an 
electromagnetic shock wave accompanied by a gravitational shock wave, each having propagation direction $\partial/\partial v$ in the space--time with line element
\begin{equation}\label{22}
ds^2=-\cos^2\{\sqrt{2}au_+\}dx^2-dy^2+2\,du\,dv\ .\end{equation}The wave amplitudes are simply related via the parameter $a$, which could be thought of as a form of 
``fine tuning". We note that the space--time is flat and the fields vanish if, in addition to $v<0$, we have $u<0$. A similar situation arises in the region of space--time corresponding 
to $u<0$ with the gravitational shock wave described by $\Psi_0=b^2\,\vartheta(v)$ and the electromagnetic shock wave described by $b\,\vartheta(v)\vartheta^3\wedge\vartheta^1$, each 
having now propagation direction $\partial/\partial u$ in the space--time with line element
\begin{equation}\label{22'}
ds^2=-\cos^2\{\sqrt{2}bv_+\}dx^2-dy^2+2\,du\,dv\ .\end{equation}The wave amplitudes are again ``fine tuned" via the parameter $b$. The electromagnetic and gravitational fields are 
non--vanishing in the region $v>0$ and vanish in the flat region $v<0$. After these two light--like signals collide at $u=v=0$ we 
obtain the post--collision region of space--time $u\geq0, v\geq0$. Clearly the subset $u>0, v>0$ is given by the Cartesian product space--time described in Section 2. This space--time 
includes a cosmological constant which has been considered in some works \cite{Pe} as a possible candidate for dark energy and appears here as a consequence of the collision. 
On the boundary $u=0, v>0$ of this region we see from (\ref{19}) that there is a light--like shell of matter with this boundary as history in space--time 
(a 2--plane of matter traveling with the speed of light, for example \cite{BH}) and from the last equation in (\ref{21}) there is 
an impulsive gravitational wave with this boundary as history in space--time. Similarly the boundary $v=0, u>0$ is the history in space--time of a light--like shell of matter following from (\ref{19}) and of 
an impulsive gravitational wave following from the first equation in (\ref{21}). These products of the collision, the light--like shells, the impulsive gravitational waves, the cosmological constant, can be 
thought of as a complicated redistribution of the energy in the incoming light--like signals. Such phenomena occur in most collisions involving thin shells, impulsive waves and shock waves, 
and are a consequence of the interactions between matter and the electromagnetic and gravitational fields \cite{BH}. Additionally one can have black hole production from the collision of two 
ultra--relativistic particles \cite{EG}, the mass inflation phenomenon inside a black hole \cite{PI}, \cite{BIP} and the production of radiation from the collision of shock waves \cite{HS}, \cite{CH}.

\end{document}